# Evaluation of the Causal Effect of Control Plans in Nonrecursive Structural Equation Models

**Manabu Kuroki**
Department of Systems Innovation
Graduate School of Engineering Science
Osaka University
Machikaneyama-cho, Toyonaka, Osaka
mkuroki@sigmath.es.osaka-u.ac.jp

**Zhihong Cai**
Department of Biostatistics
School of Public Health
Kyoto University
Yoshida, Konoe-cho, Sakyo-ku, Kyoto
cai@pbh.med.kyoto-u.ac.jp

## Abstract

When observational data is available from practical studies and a directed cyclic graph for how various variables affect each other is known based on substantive understanding of the process, we consider a problem in which a control plan of a treatment variable is conducted in order to bring a response variable close to a target value with variation reduction. We formulate an optimal control plan concerning a certain treatment variable through path coefficients in the framework of linear nonrecursive structural equation models. Based on the formulation, we clarify the properties of causal effects when conducting a control plan. The results enable us to evaluate the effect of a control plan on the variance from observational data.

## 1 INTRODUCTION

Structural equation models are widely used in practical science with the purpose to explain causal relationships and evaluate causal effects. On the basis that the causal effects are unidirectional or not, the structural equation model can be divided into recursive and nonrecursive cases. In recursive structural equation models, no two variables can be assumed as reciprocally related with each affecting the other. However, this assumption is often at odds with our understanding of the nature of the social science process. For example, in economics, the price of a good may be a function of the quantity either demanded or supplied, while these quantities themselves may be influenced by the price or the expectation of price that consumers or suppliers may have. Thus, in many cases, it is unrealistic to assume that no two variables in a model are reciprocally related. Under such situations, it is desirable to employ nonrecursive structural equation models (Berry, 1984; Richardson 1996a).

The nonrecursive structural equation model has been discussed by Haavelmo (1943), Strotz and Wold (1960) and other researchers for decades. Recently, there has been a lot of research in the field of artificial intelligence and statistics. When the functional relationships between variables can be described as a directed cyclic graph and the corresponding linear nonrecursive structural equation model, Spirtes (1995) and Richardson (1996b) provided the graphical representation of observed conditional independencies. In addition, Koster (1995) and Richardson (1997) clarified the relationship between the d-separation criterion and conditional independencies. Further, Richardson (1996c, 1996d) and Richardson and Spirtes (1996) provided causal discovery algorithms based on the observed conditional independencies.

While the causal discovery problem in linear nonrecursive structural equation models has been studied a lot (Richardson, 1996c, 1996d; Richardson and Spirtes, 1996), the evaluation problem of causal effects remains scarce. In linear recursive structural equation models, Brito and Pearl (2002), Pearl (2000) and Tian (2004) provided the graphical identifiability criteria for total effects. In addition, Cai and Kuroki (2005), Kuroki and Miyakawa (2003) and Kuroki et al. (2003) clarified how a response variable could be changed when a control plan is conducted.

On the other hand, in linear nonrecursive structural equation models, Berry (1984) and Duncan (1974) stated that the two stage least square method is useful to evaluate causal effects, which is often referred to as the instrumental variable method (Bowden and Turkington, 1984; Brito and Pearl, 2002; Duncan, 1975). However, it is not clear how a control plan should be conducted in order to bring a response variable close to a target value with variation reduction. Therefore, it is necessary to establish a new framework of linear nonrecursive structural equation models in order to evaluate the causal effects of a control plan.

When observed data is available from practical studies, this paper considers a situation where cause-effect relationships between variables can be described as a directed cyclic graph and the corresponding linear structural equation model. Then, according to Pearl (2000), we define an external intervention as an operation in which a certain structural equation is replaced by a planning formula. Based on this definition, we discuss a problem in which a control plan of a treatment variable is conducted in order to bring a response variable close to a target value with variation reduction. Different from linear recursive structural equation models where an optimal plan is usually achieved, in linear nonrecursive structural equation models, the stable condition is not always satisfied, which indicates that the variance can not be evaluated uniquely. Therefore, it is necessary to investigate the stability of a linear nonrecursive structural equation model when conducting a control plan.

In this paper, we first formulate the causal effect of a control plan through path coefficients and observed covariance. Based on this formulation, the stable condition when conducting a control plan is clarified. Next, we clarify the causal mechanism for how the mean and the variance of a response variable would change when a control plan is conducted. Finally, we illustrate our results through an empirical study. The results enable us to identify and to estimate the effect of a control plan on the variance of the response variable in the framework of linear nonrecursive structural equation models.

## 2 PRELIMINARIES

### 2.1 LINEAR STRUCTRAL EQUATION MODEL

In statistical causal analysis, a directed graph that represents cause-effect relationships is called a path diagram. A directed graph is a pair $G = (\boldsymbol{V}, \boldsymbol{E})$, where $\boldsymbol{V}$ is a finite set of vertices and the set $\boldsymbol{E}$ of arrows is a subset of the set $\boldsymbol{V} \times \boldsymbol{V}$ of ordered pairs of distinct vertices. Regarding graph theoretic terminology used in this paper, see Kuroki and Cai (2004).

**DEFINITION (PATH DIAGRAM)**

Suppose a directed graph $G = (\boldsymbol{V}, \boldsymbol{E})$ with a set $\boldsymbol{V} = \{V_1, \cdots, V_n\}$ of variables is given. The graph $G$ is called a path diagram, when each child-parent family in the graph $G$ represents a linear structural equation model

$$V_i = \mu_{v_i \cdot pa(v_i)} + \sum_{V_j \in \text{pa}(V_i)} \alpha_{v_i v_j} V_j + \epsilon_{v_i}, \ i = 1, \ldots, n, \quad (1)$$

where $\text{pa}(V_i)$ denotes a set of parents of $V_i$ in $G$ and random disturbances $\epsilon_{v_1}, \ldots, \epsilon_{v_n}$ are assumed to be independent and have mean 0. In addition, $\mu_{v_i \cdot pa(v_i)}$ is a constant value and $\alpha_{v_i v_j}(\neq 0)$ is called a path coefficient or a direct effect. □

When the directed graph includes cycles, the corresponding structural equation model is said to be nonrecursive, otherwise it is said to be recursive. For detailed discussion regarding linear structural equation models, refer to Bollen (1989) and Duncan (1975).

Here, in order to continue our discussion, let us define some notations. $\mu_y$ and $\boldsymbol{\mu}_w$ represent the mean of $Y$ and the mean vector of $\boldsymbol{W}$ respectively. Let $\sigma_{xx \cdot z}$ and $\sigma_{xy \cdot z}$ be the conditional variance $\text{var}(X|\boldsymbol{Z})$ of $X$ given $\boldsymbol{Z}$ and the conditional covariance $\text{cov}(X, Y|\boldsymbol{Z})$ between $X$ and $Y$ given $\boldsymbol{Z}$, respectively. In addition, the regression coefficient $\beta_{yx \cdot z}$ of $x$ in the regression model of $Y$ on $x$ and $\boldsymbol{z}$ is defined as $\beta_{yx \cdot z} = \sigma_{xy \cdot z}/\sigma_{xx \cdot z}$. Furthermore, let $\Sigma_{ww \cdot z}$ and $\Sigma_{sw \cdot z}$ be the conditional covariance matrix of $\boldsymbol{W}$ given $\boldsymbol{Z}$ and the conditional covariance matrix between $\boldsymbol{S}$ and $\boldsymbol{W}$ given $\boldsymbol{Z}$, respectively. We let $B_{sw \cdot z} = \Sigma_{sw \cdot z} \Sigma_{ww \cdot z}^{-1}$. When $\boldsymbol{Z}$ is an empty set, $\boldsymbol{Z}$ is omitted from the notations above. Similar notations are used for other parameters.

### 2.2 STABLE CONDITION

Letting $A$ be a path coefficient matrix $A = (\alpha_{v_i v_j})$ in equation (1), linear structural equation model (1) can be rewritten as

$$\boldsymbol{V} = \boldsymbol{\mu}_{v \cdot pa(v)} + A\boldsymbol{V} + \boldsymbol{\epsilon}_v, \quad (2)$$

where $\boldsymbol{\epsilon}_v = (\epsilon_{v_1}, \cdots, \epsilon_{v_n})'$ and $\boldsymbol{\mu}_{v \cdot pa(v)} = (\mu_{v_1 \cdot pa(v_1)}, \cdots, \mu_{v_n \cdot pa(v_n)})'$. Here, denoting $n_v$ as the number of elements in $\boldsymbol{V}$, $I_{n_v, n_v}$ represents an $n_v$ dimensional identity matrix. Similar notations are used for other parameters. There are many representations equivalent to equation (2). Letting $A^0 = I_{n_v, n_v}$, substituting the right hand side of equation (2) for $\boldsymbol{V}$ in the right hand side yields

$$\begin{aligned} \boldsymbol{V} &= \boldsymbol{\mu}_{v \cdot pa(v)} + A\boldsymbol{V} + \boldsymbol{\epsilon}_v \\ &= (I_{n_v, n_v} + A)(\boldsymbol{\mu}_{v \cdot pa(v)} + \boldsymbol{\epsilon}_v) + A^2 \boldsymbol{V}, \end{aligned}$$

which is true if equation (2) is true. Performing this operation $k$ times yields

$$\boldsymbol{V} = \sum_{i=0}^{k-1} A^i (\boldsymbol{\mu}_{v \cdot pa(v)} + \boldsymbol{\epsilon}_v) + A^k \boldsymbol{V}.$$

When both $A^k$ and $\sum_{i=0}^{k-1} A^i$ converge to specific matrices respectively, the linear structural equation model is said to be stable (Bentler and Freeman, 1983).

Here, a matrix $A$ is said to be convergent (Ben-Israel and Greville, 1972) if the following equation holds true:

$$\lim_{k \to 0} A^k = 0.$$

It is known that a matrix $A$ is convergent if and only if the absolute value of the largest eigen value is less than one (Bentler and Freeman, 1983). In addition, Bentler and Freeman (1983) stated that

$$(I_{n_v, n_v} - A)^{-1} = \sum_{k=0}^{\infty} A^k$$

holds true if $A$ is a convergent matrix.

Stability indicates a situation that the observed data achieves a steady state. Under such a situation, it is possible to consider conducting a control plan, since the mean and the variance can be evaluated uniquely. Thus, the stable condition must be satisfied in order to evaluate the causal effects on the mean and the variance.

## 3 CONTROL PLAN

### 3.1 PROBLEM DESCRIPTION

In linear nonrecursive structural equation models, since two variables are reciprocally causative, it may be difficult to tell which one is the cause and which one is the effect. However, in practical science, a policy maker would like to know how to influence a variable $Y$ by manipulating a variable $X$. In this case, we call $X$ a treatment variable, and $Y$ a response variable. When observational data are available and the cause-effect relationships between variables can be described as a directed cyclic graph shown in Fig. 1, we consider a problem in which a control plan of a treatment variable is conducted in order to bring a response variable close to a target value with variation reduction.

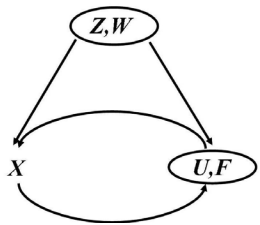

Fig. 1: Problem description

In Fig. 1, $X$ stands for a treatment variable, which can be manipulated. Both $U$ and $F$ are sets of variables which are affected by $X$ and may have an effect on $X$. Here, $F$ includes $Y$ which is a response variable of interest. In addition, $Z$ and $W$ represent sets of covariates, which can be determined before assigning values to $X$.

Then, in order to bring the value of $Y$ close to a target value, consider conducting a control plan in which $X$ is set to be the following linear function:

$$X = x + \boldsymbol{a}'\boldsymbol{F} + \boldsymbol{b}'\boldsymbol{W} + \epsilon_x^* = h(\boldsymbol{F}, \boldsymbol{W}, \epsilon_x^*), \quad (3)$$

where $x$ is a constant value and both $\boldsymbol{a}$ and $\boldsymbol{b}$ are constant vectors. In addition, $\epsilon_x^*$ is a random disturbance with mean 0 and variance $\sigma_{\epsilon_x^* \epsilon_x^*}$ when conducting a control plan, and it is assumed to be independent of the other random disturbances. Furthermore, both $\boldsymbol{F}$ and $\boldsymbol{W}$ are called variables for control in this paper, since these variables are used for conducting a control plan of $X$.

If $\boldsymbol{a}$ is a nonzero vector, equation (3) is called a nonrecursive control plan, otherwise it is called a recursive control plan. In addition, when $\sigma_{\epsilon_x^* \epsilon_x^*}$ is equal to zero, equation (3) is called a perfect control plan, otherwise it is called an imperfect control plan. An imperfect plan indicates that the treatment variable could not be manipulated exactly due to random disturbance. It is important to evaluate causal effects when conducting an imperfect control plan, since we can not always achieve a perfect control plan in practical science.

In order to clarify how the response variable would have changed if the control plan (3) were conducted, we formulate both the mean $E(Y|\text{set}(X = h(\boldsymbol{F}, \boldsymbol{W}, \epsilon_x^*)))$ and the variance $\text{var}(Y|\text{set}(X = h(\boldsymbol{F}, \boldsymbol{W}, \epsilon_x^*)))$ through the path coefficients. Here, $\text{set}(X = h(\boldsymbol{F}, \boldsymbol{W}, \epsilon_x^*))$ means that we set an original equation of $X$ to $X = h(\boldsymbol{F}, \boldsymbol{W}, \epsilon_x^*)$ by an external intervention. In this paper, the mean and the variance when conducting a control plan are called "the causal effects on the mean" and "the causal effect on the variance", respectively. As seen from section 3.2, in the case of linear nonrecursive structural equation models, the causal effect does not always exist, which is dependent on the planning formula. The stable condition must be satisfied in order to evaluate causal effects when conducting a control plan.

### 3.2 FORMULATION

According to section 3.1, we partition a set $\boldsymbol{V}$ of vertices in a path diagram $G$ into the following three disjoint sets:

$\boldsymbol{S} = \boldsymbol{F} \cup \boldsymbol{U}$: a set of descendants of $X$ ($Y \in \boldsymbol{F}$, $\boldsymbol{F} \cap \boldsymbol{U} = \phi$). Here, $\boldsymbol{F}$ includes the first $n_f$ components of $\boldsymbol{S}$. In addition, $Y$, which is a response variable of interest, is the first component of $\boldsymbol{F}$,

$X$ : a treatment variable,

$\boldsymbol{T} = \boldsymbol{W} \cup \boldsymbol{Z} = \boldsymbol{V} \setminus (\{X\} \cup \boldsymbol{S})$ : a set of nondescendants of $X$ ($\boldsymbol{W} \cap \boldsymbol{Z} = \phi$). Here, $\boldsymbol{W}$ consists of the first $n_w$ components of $\boldsymbol{T}$.

According to the above partition of $\boldsymbol{V}$, let $A_{st}$ be a path coefficient matrix of $\boldsymbol{T}$ on $\boldsymbol{S}$ whose $(i,j)$ component is the path coefficient of $T_j$ on $S_i$ ($S_i \in \boldsymbol{S}, T_j \in \boldsymbol{T}$). Let $\boldsymbol{0}_{n_x,n_s}$ be an $(n_x, n_s)$ zero matrix. Similar notations are used for other matrices.

Then, equation (2) can be rewritten as follows:

$$\begin{pmatrix} \boldsymbol{S} \\ X \\ \boldsymbol{T} \end{pmatrix} = \begin{pmatrix} \boldsymbol{\mu}_{s\cdot pa(s)} \\ \mu_{x\cdot pa(x)} \\ \boldsymbol{\mu}_{t\cdot pa(t)} \end{pmatrix} + \begin{pmatrix} \boldsymbol{\epsilon}_s \\ \epsilon_x \\ \boldsymbol{\epsilon}_t \end{pmatrix}$$
$$+ \begin{pmatrix} A_{ss} & A_{sx} & A_{st} \\ A_{xs} & 0 & A_{xt} \\ \boldsymbol{0}_{n_t,n_s} & \boldsymbol{0}_{n_t,n_x} & A_{tt} \end{pmatrix} \begin{pmatrix} \boldsymbol{S} \\ X \\ \boldsymbol{T} \end{pmatrix}, \quad (4)$$

where $\boldsymbol{\epsilon}_s$ and $\boldsymbol{\epsilon}_t$ are random disturbance vectors corresponding to $\boldsymbol{S}$ and $\boldsymbol{T}$, respectively. In addition, $\boldsymbol{\mu}_{s\cdot pa(s)}$ and $\boldsymbol{\mu}_{t\cdot pa(t)}$ are constant vectors corresponding to $\boldsymbol{S}$ and $\boldsymbol{T}$, respectively. Similar notations are used for other vectors.

Then, letting

$$A = \begin{pmatrix} A_{ss} & A_{sx} & A_{st} \\ A_{xs} & 0 & A_{xt} \\ \boldsymbol{0}_{ts} & \boldsymbol{0}_{tx} & A_{tt} \end{pmatrix}, A_{11} = \begin{pmatrix} A_{ss} & A_{sx} \\ A_{xs} & 0 \end{pmatrix},$$

the characteristic equation of $A$ can be provided as

$$\det(\lambda I_{n_v,n_v} - A)$$
$$= \det(\lambda I_{n_t,n_t} - A_{tt})\det(\lambda I_{n_s+1,n_s+1} - A_{11}) = 0.$$

Thus, both $A_{tt}$ and $A_{11}$ must be convergent matrices in order that equation (4) satisfies the stable condition. Here, letting $C_{xs} = (\boldsymbol{a}'; \boldsymbol{0}_{1,n_u})$ and $C_{xt} = (\boldsymbol{b}'; \boldsymbol{0}_{1,n_z})$, the linear structural equation model after conducting a control plan can be represented as

$$\begin{pmatrix} \boldsymbol{S} \\ X \\ \boldsymbol{T} \end{pmatrix} = \begin{pmatrix} \boldsymbol{\mu}_{s\cdot pa(s)} \\ x \\ \boldsymbol{\mu}_{t\cdot pa(t)} \end{pmatrix} + \begin{pmatrix} \boldsymbol{\epsilon}_s \\ \epsilon_x^* \\ \boldsymbol{\epsilon}_t \end{pmatrix}$$
$$+ \begin{pmatrix} A_{ss} & A_{sx} & A_{st} \\ C_{xs} & 0 & C_{xt} \\ \boldsymbol{0}_{nz,n_t} & \boldsymbol{0}_{nt,n_x} & A_{tt} \end{pmatrix} \begin{pmatrix} \boldsymbol{S} \\ X \\ \boldsymbol{T} \end{pmatrix}. \quad (5)$$

Similarly, in order that equation (5) satisfies the stable condition,

$$A_{11}^* = \begin{pmatrix} A_{ss} & A_{sx} \\ C_{xs} & 0 \end{pmatrix}$$

must be a convergent matrix.

In order to obtain the main result of this paper, we need the following lemma:

**LEMMA 1(Rao, 1973)**

For matrices $A$, $B$, $C$ and $D$,

$$(A - BD^{-1}C)^{-1} = A^{-1} + A^{-1}B(D - CA^{-1}B)^{-1}CA^{-1}.$$

holds true.  □

Then, based on the setting above, the following theorem can be obtained:

**THEOREM 1**

In a stable linear structural equation model, the causal effect of the control plan $\text{set}(X = h(\boldsymbol{F}, \boldsymbol{W}, \epsilon_x^*))$ on the variance of $Y$ is minimized when $\boldsymbol{b}$ satisfies

$$\boldsymbol{b} = \frac{\boldsymbol{\gamma}_{fx}'(\boldsymbol{\gamma}_{fx}B_{xw} - B_{fw})}{\boldsymbol{\gamma}_{fx}'\boldsymbol{\gamma}_{fx}} \quad (6)$$

for a given $\boldsymbol{a}$ satisfying $|\boldsymbol{a}'\boldsymbol{\gamma}_{fx}| < 1$. Here, $\boldsymbol{\gamma}_{fx}$ is a vector which is provided as the first $n_f$ components of

$$\boldsymbol{\tau}_{sx} = (I_{n_s,n_s} - A_{ss})^{-1}A_{sx}.$$

Letting $\boldsymbol{b}^*$ be equation (6) and the corresponding control plan be $\text{set}(X = g(\boldsymbol{F}, \boldsymbol{W}, \epsilon_x^*))$, the causal effect of the control plan $\text{set}(X = g(\boldsymbol{F}, \boldsymbol{W}, \epsilon_x^*))$ on the mean of $Y$ is given by the formula

$$E(Y|\text{set}(X = g(\boldsymbol{F}, \boldsymbol{W}, \epsilon_x^*))) = \gamma_{yx}(x - \mu_x + \boldsymbol{b}^{*'}\boldsymbol{\mu}_w)$$
$$+ \mu_y + \frac{\gamma_{yx}}{1 - \boldsymbol{a}'\boldsymbol{\gamma}_{fx}}\boldsymbol{a}'(\boldsymbol{\mu}_f + \boldsymbol{\gamma}_{fx}(x - \mu_x + \boldsymbol{b}^{*'}\boldsymbol{\mu}_w)),$$

where $\gamma_{yx}$ is given by the first component of $\boldsymbol{\tau}_{sx}$. Then, the causal effect of the control plan $\text{set}(X = g(\boldsymbol{F}, \boldsymbol{W}, \epsilon_x^*))$ on the variance of $Y$ is given by the $(1,1)$ component of the formula

$$\text{var}(\boldsymbol{F}|\text{set}(X = g(\boldsymbol{F}, \boldsymbol{W}, \epsilon_x^*)))$$
$$= \left(I_{n_f,n_f} + \frac{\boldsymbol{\gamma}_{fx}\boldsymbol{a}'}{1 - \boldsymbol{a}'\boldsymbol{\gamma}_{fx}}\right)\left(\Sigma_{ff} + \boldsymbol{\gamma}_{fx}\boldsymbol{\gamma}_{fx}'\sigma_{\epsilon_x^*\epsilon_x^*}\right.$$
$$- (\boldsymbol{\gamma}_{fx} - B_{fx})(\boldsymbol{\gamma}_{fx} - B_{fx})'\sigma_{xx} - B_{fx}B_{fx}'\sigma_{xx}$$
$$\left. - (B_{fw} - \boldsymbol{\gamma}_{fx}B_{xw})\Sigma_{ww}(B_{fw} - \boldsymbol{\gamma}_{fx}B_{xw})'\right)$$
$$\times \left(I_{n_f,n_f} + \frac{\boldsymbol{\gamma}_{fx}\boldsymbol{a}'}{1 - \boldsymbol{a}'\boldsymbol{\gamma}_{fx}}\right)'.$$

□

$X = g(\boldsymbol{F}, \boldsymbol{W}, \epsilon_x^*)$ in Theorem 1 is called an optimal plan of $X$ for a given $\boldsymbol{a}$.

**PROOF: THE MEAN**

Noting that equation (4) is stable, $(I_{n_s+1,n_s+1} - A_{11})^{-1}$ can be provided as

$$(I_{n_s+1,n_s+1} - A_{11})^{-1} = \begin{pmatrix} A^{ss} & A^{sx} \\ A^{xs} & A^{xx} \end{pmatrix},$$

where

$$A^{xx} = (1 - A_{xs}(I_{n_s+1,n_s+1} - A_{ss})^{-1}A_{sx})^{-1},$$
$$A^{ss} = (I_{n_s,n_s} - A_{ss})^{-1} + (I_{n_s,n_s} - A_{ss})^{-1}A_{sx}$$
$$\times A^{xx}A_{xs}(I_{n_s,n_s} - A_{ss})^{-1},$$
$$A^{sx} = (I_{n_s,n_s} - A_{ss})^{-1}A_{sx}A^{xx},$$
$$A^{xs} = A^{xx}A_{xs}(I_{n_s,n_s} - A_{ss})^{-1}.$$

This indicates that $(I_{n_s,n_s} - A_{ss})^{-1}$ also exists. Thus, from equation (4), since we can obtain

$$S = \mu_{s \cdot pa(s)} + A_{ss}S + A_{sx}X + A_{st}T + \epsilon_s, \quad (7)$$

the mean vector $\mu_s$ of $S$ can be expressed as

$$\mu_s = (I_{n_sn_s} - A_{ss})^{-1}\mu_{s \cdot pa(s)} + \tau_{sx}\mu_x + \tau_{st}\mu_t, \quad (8)$$

where $\tau_{sx} = (I_{n_sn_s} - A_{ss})^{-1}A_{sx}$ and $\tau_{st} = (I_{n_sn_s} - A_{ss})^{-1}A_{st}$.

On the other hand, by noting that the renewal structural equation model (5) is also stable, the inverse matrix of $(I_{n_s+1,n_s+1} - A_{11}^*)$ can be provided as

$$(I_{n_s+1,n_s+1} - A_{11}^*)^{-1} = \begin{pmatrix} A^{*ss} & A^{*sx} \\ A^{*xs} & A^{*xx} \end{pmatrix},$$

where

$$\begin{aligned}
A^{*xx} &= (1 - C_{xs}(I_{n_s+1,n_s+1} - A_{ss})^{-1}A_{sx})^{-1}, \\
A^{*ss} &= (I_{n_s,n_s} - A_{ss} - A_{sx}C_{xs})^{-1} \\
A^{*sx} &= (I_{n_s,n_s} - A_{ss})^{-1}A_{sx}A^{*xx}, \\
A^{*xs} &= A^{*xx}C_{xs}(I_{n_s,n_s} - A_{ss})^{-1}.
\end{aligned}$$

This indicates that $(I_{n_s,n_s} - A_{ss} - A_{xs}C_{sx})^{-1}$ also exists. Thus, since we can obtain from equation (5)

$$\begin{aligned}
S &= \mu_{s \cdot pa(s)} + A_{ss}S + A_{st}T + \epsilon_s \\
&\quad + A_{sx}(x + C_{xs}S + C_{xt}T + \epsilon_x^*) \\
&= \mu_{s \cdot pa(s)} + A_{sx}(x + \epsilon_x^*) + \epsilon_s \\
&\quad + (A_{ss} + A_{sx}C_{xs})S + (A_{st} + A_{sx}C_{xt})T, \quad (9)
\end{aligned}$$

by noting that $(I_{n_s,n_s} - A_{ss})^{-1}$ exists, we can derive

$$\begin{aligned}
&E(S|\text{set}(X = h(F, W, \epsilon_x^*))) \\
&= (I_{n_s,n_s} - A_{ss} - A_{sx}C_{xs})^{-1}\{\mu_{s \cdot pa(s)} \\
&\quad + A_{sx}x + (A_{st} + A_{sx}C_{xt})\mu_t\} \\
&= (I_{n_s,n_s} - \tau_{sx}C_{xs})^{-1}(\mu_s + \tau_{sx}(x - \mu_x + C_{xt}\mu_t))
\end{aligned}$$

from equation (8). Here, by noting $C_{xs} = (a'; 0_{1,n_u})$ and $C_{xt} = (b'; 0'_{1,n_z})$, since $A_{sx}C_{xs}$ is a matrix with rank one if it is a non-zero matrix, it has only one non-zero eigen value $C_{xs}\tau_{sx} = a'\gamma_{fx}$, which must satisfy $|a'\gamma_{fx}| < 1$ in order to obtain the stable structural equation model. In the case where $A_{sx}C_{xs}$ is a zero matrix, every eigen value is zero, which is also satisfied the stable condition $|a'\gamma_{fx}| < 1$.

From Lemma 1, we can obtain

$$\begin{aligned}
(I_{n_s,n_s} - \tau_{sx}C_{xs})^{-1} &= I_{n_s,n_s} + \frac{\tau_{sx}C_{xs}}{1 - a'\gamma_{fx}}, \\
&= \begin{pmatrix} D_1 & 0_{n_f,f_u} \\ D_2 & I_{n_u,n_u} \end{pmatrix},
\end{aligned}$$

where

$$D_1 = I_{n_f,n_f} + \frac{\gamma_{fx}a'}{1 - a'\gamma_{fx}}, \quad D_2 = \frac{\gamma_{ux}a'}{1 - a'\gamma_{fx}}.$$

In addition, $\gamma_{ux}$ is the last $n_u$ components of $\tau_{sx}$. Thus, we can derive

$$\begin{aligned}
&E(S|\text{set}(X = h(F, W, \epsilon_x^*))) \\
&= (I_{n_s,n_s} + \frac{\tau_{sx}C_{xs}}{1 - a'\gamma_{fx}})(\mu_s + \tau_{sx}(x - \mu_x + b'\mu_w)).
\end{aligned}$$

Therefore,

$$\begin{aligned}
E(Y|\text{set}(X = h(F, W, \epsilon_x^*))) &= \gamma_{yx}(x - \mu_x + b'\mu_w) \\
+\mu_y + \frac{\gamma_{yx}}{1 - a'\gamma_{fx}}a'(\mu_f &+ \gamma_{fx}(x - \mu_x + b'\mu_w)).
\end{aligned}$$

**PROOF: THE VARIANCE**

Regarding the variance of $\Sigma_{\epsilon_s\epsilon_s}$, we can obtain

$$\Sigma_{st} = \tau_{sx}\Sigma_{xt} + \tau_{st}\Sigma_{tt},$$

from equation (7). In addition, since equation (7) can be rewritten as

$$\begin{aligned}
S - \tau_{sx}X &= (I_{n_s,n_s} - A_{ss})^{-1}\mu_{s \cdot pa(s)} \\
&\quad + \tau_{st}T + (I_{n_s,n_s} - A_{ss})^{-1}\epsilon_s,
\end{aligned}$$

we can obtain

$$\begin{aligned}
&(I_{n_s,n_s} - A_{ss})^{-1}\Sigma_{\epsilon_s\epsilon_s}(I_{n_s,n_s} - A_{ss})'^{-1} \\
&= \Sigma_{ss} - \tau_{sx}\Sigma_{xs} - \Sigma_{sx}\tau'_{sx} + \tau_{sx}\tau'_{sx}\sigma_{xx} \\
&\quad - \tau_{st}\Sigma_{tt}\tau'_{st} \\
&= \Sigma_{ss} + (\tau_{sx} - B_{sx})(\tau_{sx} - B_{sx})'\sigma_{xx} \\
&\quad - B_{sx}B'_{sx}\sigma_{xx} - \tau_{st}\Sigma_{tt}\tau'_{st}.
\end{aligned}$$

Furthermore, by noting $(\tau_{st} + \tau_{sx}C_{xt}) = (\tau_{sw} + \tau_{sx}b'; \tau_{sz})$, from equation (9), we can obtain

$$\begin{aligned}
&(I_{n_s,n_s} - \tau_{sx}C_{xs})\text{var}(S|\text{set}(X = h(Y, W, \epsilon_x^*))) \\
&\quad \times (I_{n_s,n_s} - \tau_{sx}C_{xs})' \\
&= (\tau_{st} + \tau_{sx}C_{xt})\Sigma_{tt}(\tau_{st} + \tau_{sx}C_{xt})' + \tau_{sx}\tau'_{sx}\sigma_{\epsilon_x^*\epsilon_x^*} \\
&\quad + (I_{n_s,n_s} - A_{ss})^{-1}\Sigma_{\epsilon_s\epsilon_s}(I_{n_s,n_s} - A_{ss})'^{-1} \\
&= (\tau_{sw} + \tau_{sx}b', \tau_{sz})\begin{pmatrix} \Sigma_{ww} & \Sigma_{wz} \\ \Sigma_{zw} & \Sigma_{zz} \end{pmatrix} \\
&\quad \times \begin{pmatrix} \tau'_{sw} + b\tau'_{sx} \\ \tau'_{sz} \end{pmatrix} + \Sigma_{ss} - B_{sx}B'_{sx}\sigma_{xx} \\
&\quad + (\tau_{sx} - B_{sx})(\tau_{sx} - B_{sx})'\sigma_{xx} + \tau_{sx}\tau'_{sx}\sigma_{\epsilon_x^*\epsilon_x^*} \\
&\quad - (\tau_{sw}, \tau_{sz})\begin{pmatrix} \Sigma_{ww} & \Sigma_{wz} \\ \Sigma_{zw} & \Sigma_{zz} \end{pmatrix}\begin{pmatrix} \tau'_{sw} \\ \tau'_{sz} \end{pmatrix} \\
&= \Sigma_{ss} - B_{sx}B'_{sx}\sigma_{xx} + (\tau_{sx} - B_{sx})(\tau_{sx} - B_{sx})'\sigma_{xx} \\
&\quad + (\tau_{sw} + \tau_{sx}b')\Sigma_{ww}(\tau_{sw} + \tau_{sx}b')' + \tau_{sx}\tau'_{sx}\sigma_{\epsilon_x^*\epsilon_x^*}
\end{aligned}$$

$$+(\boldsymbol{\tau}_{sw} + \boldsymbol{\tau}_{sx}\boldsymbol{b}')\Sigma_{wz}\boldsymbol{\tau}'_{sz} + \boldsymbol{\tau}_{sz}\Sigma_{zw}(\boldsymbol{\tau}_{sw} + \boldsymbol{\tau}_{sx}\boldsymbol{b}')'$$
$$+\boldsymbol{\tau}_{sz}\Sigma_{zz}\boldsymbol{\tau}'_{sz} - \boldsymbol{\tau}_{sw}\Sigma_{ww}\boldsymbol{\tau}'_{sw}$$
$$-\boldsymbol{\tau}_{sw}\Sigma_{wz}\boldsymbol{\tau}'_{sz} - \boldsymbol{\tau}_{sz}\Sigma_{zw}\boldsymbol{\tau}'_{sw} - \boldsymbol{\tau}_{sz}\Sigma_{zz}\boldsymbol{\tau}'_{sz}$$
$$= \Sigma_{ss} - B_{sx}B'_{sx}\sigma_{xx} + (\boldsymbol{\tau}_{sx} - B_{sx})(\boldsymbol{\tau}_{sx} - B_{sx})'\sigma_{xx}$$
$$+\boldsymbol{\tau}_{sx}\boldsymbol{\tau}'_{sx}\sigma_{\epsilon^*_x\epsilon^*_x} + (\boldsymbol{\tau}_{sw} + \boldsymbol{\tau}_{sx}\boldsymbol{b}' + \boldsymbol{\tau}_{sz}B_{zw})\Sigma_{ww}$$
$$\times(\boldsymbol{\tau}_{sw} + \boldsymbol{\tau}_{sx}\boldsymbol{b}' + \boldsymbol{\tau}_{sz}B_{zw})'$$
$$-(\boldsymbol{\tau}_{sw} + \boldsymbol{\tau}_{sz}B_{zw})\Sigma_{ww}(\boldsymbol{\tau}_{sw} + \boldsymbol{\tau}_{sz}B_{zw})'.$$

Here, by noting

$$\Sigma_{sw} = \boldsymbol{\tau}_{sx}\Sigma_{xw} + \boldsymbol{\tau}_{sz}\Sigma_{zw} + \boldsymbol{\tau}_{sw}\Sigma_{ww}$$

from equation (7), we can obtain

$$(I_{n_s,n_s} - \boldsymbol{\tau}_{sx}C_{xs})\text{var}(\boldsymbol{S}|\text{set}(X = h(Y, \boldsymbol{W}, \epsilon^*_x)))$$
$$\times(I_{n_s,n_s} - \boldsymbol{\tau}_{sx}C_{xs})'$$
$$= \Sigma_{ss} - B_{sx}B'_{sx}\sigma_{xx} + (\boldsymbol{\tau}_{sx} - B_{sx})(\boldsymbol{\tau}_{sx} - B_{sx})'\sigma_{xx}$$
$$+\boldsymbol{\tau}_{sx}\boldsymbol{\tau}'_{sx}\sigma_{\epsilon^*_x\epsilon^*_x} + (\boldsymbol{\tau}_{sx}\boldsymbol{b}' + B_{sw} - \boldsymbol{\tau}_{sx}B_{xw})\Sigma_{ww}$$
$$\times(\boldsymbol{\tau}_{sx}\boldsymbol{b}' + B_{sw} - \boldsymbol{\tau}_{sx}B_{xw})'$$
$$-(B_{sw} - \boldsymbol{\tau}_{sx}B_{xw})\Sigma_{ww}(B_{sw} - \boldsymbol{\tau}_{sx}B_{xw})'.$$

This equation is not dependent on $\boldsymbol{U}$ or $\boldsymbol{Z}$. Then, since the first $n_s$ rows of $(I_{n_s,n_s} - A_{sx}C_{xs})^{-1}$ can be provided as

$$\left(I_{n_f,n_f} + \frac{\boldsymbol{\gamma}_{fx}\boldsymbol{a}'}{1 - \boldsymbol{a}'\boldsymbol{\gamma}_{fx}}, \boldsymbol{0}_{n_f,n_u}\right),$$

in order to minimize the variance of $Y$ for a given $\boldsymbol{a}$ satisfying $|\boldsymbol{a}'\boldsymbol{\tau}_{sx}| < 1$, we can solve the following equation regarding $\boldsymbol{b}$:

$$\boldsymbol{\gamma}_{fw} + \boldsymbol{\gamma}_{fx}\boldsymbol{b}' + \boldsymbol{\gamma}_{fz}B_{zw}$$
$$= \boldsymbol{\gamma}_{fx}\boldsymbol{b}' + B_{fw} - \boldsymbol{\gamma}_{fx}B_{xw} = \boldsymbol{0}_{n_f,n_w}. \quad (10)$$

When $\boldsymbol{\gamma}_{fx}$ is a nonzero vector, the solution of equation (10) can be given as

$$\boldsymbol{b}' = \frac{\boldsymbol{\gamma}'_{fx}(\boldsymbol{\gamma}_{fx}B_{xw} - B_{fw})}{\boldsymbol{\gamma}'_{fx}\boldsymbol{\gamma}_{fx}}.$$

This equation is dependent on the selection of $\boldsymbol{W} \cup \boldsymbol{F}$ but not on $\boldsymbol{Z} \cup \boldsymbol{U}$. Letting $\boldsymbol{b}^*$ be this equation and the corresponding control plan be $\text{set}(X = g(\boldsymbol{F}, \boldsymbol{W}, \epsilon^*_x))$,

$$\text{var}(\boldsymbol{F}|\text{set}(X = g(\boldsymbol{F}, \boldsymbol{W}, \epsilon^*_x)))$$
$$= \left(I_{n_f,n_f} + \frac{\boldsymbol{\gamma}_{fx}\boldsymbol{a}'}{1 - \boldsymbol{a}'\boldsymbol{\gamma}_{fx}}\right)\left(\Sigma_{ff} + \boldsymbol{\gamma}_{fx}\boldsymbol{\gamma}'_{fx}\sigma_{\epsilon^*_x\epsilon^*_x}\right.$$
$$+(\boldsymbol{\gamma}_{fx} - B_{fx})(\boldsymbol{\gamma}_{fx} - B_{fx})'\sigma_{xx} - B_{fx}B'_{fx}\sigma_{xx}$$
$$\left.-(B_{fw} - \boldsymbol{\gamma}_{fx}B_{xw})\Sigma_{ww}(B_{fw} - \boldsymbol{\gamma}_{fx}B_{xw})'\right)$$
$$\times \left(I_{n_f,n_f} + \frac{\boldsymbol{\gamma}_{fx}\boldsymbol{a}'}{1 - \boldsymbol{a}'\boldsymbol{\gamma}_{fx}}\right)'$$

This equation is dependent on the selection of $\boldsymbol{W} \cup \boldsymbol{U}$ but not on $\boldsymbol{Z} \cup \boldsymbol{U}$.                              □

We would like to point out some properties of our formulation. Since Theorem 1 is based on linear structural equation models, we can apply the identifiability criteria for total effects of $X$ on $\boldsymbol{F}$ in linear structural equation model to evaluate the causal effect on the variance. For example, for a given $\sigma^*_{xx}$ and $\boldsymbol{a}$, the two stage least square method for the $\boldsymbol{\gamma}_{fx}$ can be used to identify the causal effect on the mean and the variance. In addition, for two optimal control plans $X = x + \boldsymbol{a}'\boldsymbol{F} + \boldsymbol{b}_1^{*'}\boldsymbol{W}_1 + \epsilon^*_x$ and $X = x + \boldsymbol{a}'\boldsymbol{F} + \boldsymbol{b}_2^{*'}\boldsymbol{W}_2 + \epsilon^*_x$ ($Y \in \boldsymbol{F}$),

$$\text{Var}(Y|\text{set}(X = x + \boldsymbol{a}'\boldsymbol{F} + \boldsymbol{b}_1^{*'}\boldsymbol{W}_1 + \epsilon^*_x))$$
$$\leq \text{Var}(Y|\text{set}(X = x + \boldsymbol{a}'\boldsymbol{F} + \boldsymbol{b}_2^{*'}\boldsymbol{W}_2 + \epsilon^*_x))$$

holds true if

$$(B_{fw_1} - \boldsymbol{\gamma}_{fx}B_{xw_1})\Sigma_{w_1w_1}(B_{fw_1} - \boldsymbol{\gamma}_{fx}B_{xw_1})$$
$$-(B_{fw_2} - \boldsymbol{\gamma}_{fx}B_{xw_2})\Sigma_{w_2w_2}(B_{fw_2} - \boldsymbol{\gamma}_{fx}B_{xw_2})$$

is a positive semidefinite matrix, which provides a covariate selection criterion to obtain smaller variance for a given $\boldsymbol{a}$. Furthermore, as seen from equation (10), the covariance matrix between $Y$ and $\boldsymbol{W}$ is a zero matrix after conducting a control plan, which indicates that an optimal plan is one that makes the covariances between $Y$ and $\boldsymbol{W}$ to become zeros.

## 4    EXAMPLE

We illustrate our results through an empirical study reported by Iverson et al. (1984). This study is to investigate the impact of student-faculty contact on the educational aspiration level of commuter college freshmen. The size of the sample is 213 and the variables of interest in this paper are the following:

$X$: student-faculty contact,

$Y$: postfreshman year level of educational aspirations,

$Z_1$: preenrollment variables, which include high school achievement, academic aptitude, etc.,

$Z_2$: other college experiences, which include college grade-point average, peer group relations impact, etc.,

$Z_3$: faculty relations, which include interaction with faculty, faculty concern for student development, etc.

According to Iverson et al. (1984), the common assumption is that student-faculty contact "causes" postfreshman year level of educational aspirations. On the other hand, it is possible that causality flows from aspiration to contact. That is, students with initially high aspiration may seek out contact with faculty as a way of realizing their goals. Then, a loop between contact and aspiration was proposed in which changes in either variable would produce a feedback on the other. On the basis of these considerations, we consider $X$ as a treatment variable and $Y$ as a response variable in this model.

Iverson et al. (1984) provided the directed cyclic graph shown in Fig. 2, but they did not give the covariance matrix [1]. Hence, we calculate the covariance matrix based on Table 5 in Iverson et al. (1984), which is shown in Table 1. Here, it is assumed that all the variables have mean 0.

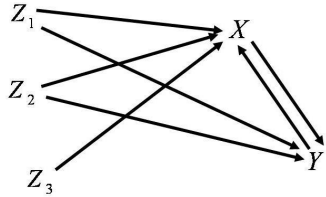

Fig. 2: Directed cyclic graph model (Iversion et al., 1984).

Table 1. The covariance matrix

|       | Y      | X      | $Z_1$  | $Z_2$  | $Z_3$  |
|-------|--------|--------|--------|--------|--------|
| Y     | 1.041  | 0.386  | -0.085 | 0.008  | 0.003  |
| X     | 0.386  | 1.216  | -0.295 | -0.038 | 0.061  |
| $Z_1$ | -0.085 | -0.295 | 1.000  | 0.000  | 0.000  |
| $Z_2$ | 0.008  | -0.038 | 0.000  | 1.000  | 0.000  |
| $Z_3$ | 0.003  | 0.061  | 0.000  | 0.000  | 1.000  |

As seen from section 3.2, in order to evaluate the causal effect on the mean and the variance of $X$ on $Y$, it is necessary to evaluate the total effect $\gamma_{yx}$ of $X$ on $Y$. The instrumental variable (IV) method (Bowden and Turkington, 1984; Brito and Pearl, 2002; Duncan, 1975) is well known as one of statistical methods to evaluate total effects in nonrecursive linear structural equation models. Since $Z_3$ is an IV regarding to $(X, Y)$, the total effect $\gamma_{yx}$ of $X$ on $Y$ is given as $\sigma_{yz_3}/\sigma_{xz_3} = 0.049$ by using the IV method.

Here, we consider conducting an unconditional plan $set(X = x)$ of $X$. Then, since there is no cycle in the graph obtained by deleting the arrow pointing from $Y$ to $X$ in Fig. 2, the corresponding structural equation model becomes stable by the unconditional plan. By conducting this plan, the mean changes from 0 to $E(Y|set(X = x)) = 0.049x$. In addition, the variance of $Y$ changes from $var(Y) = 1.041$ to $var(Y|set(X = x)) = 0.998$. These results indicate that if an educational plan aimed at raising the student-faculty contact is conducted, then (i) the average students' educa-

---

[1] According to personal communication with Prof. Terenzini P. T., they no longer have the covariance matrix used in the paper.

---

tional aspiration would improve, and (ii) the difference among students' aspiration would be reduced.

Next, we consider conducting a conditional plan $set(X = x + aY)$ of $X$. Different from the case of an unconditional plan, the graph of Fig. 2 has a cycle $X \to Y \to X$. Thus, in order to keep a stable situation, the value of $a$ must satisfy $|0.049a| < 1$. Under this condition, the mean and the variance of $Y$ after conducting such a plan are $E(Y|set(X = x + aY)) = 0.049/(1 - 0.049a)x$ and $var(Y|set(X = x + aY)) = 0.998/(1 - 0.049a)^2$, respectively. Here, the closer $a$ is to $-1/0.049$ by conducting the conditional plan, the closer the mean and the variance are to $0.049/2x = 0.025x$ and $0.998/4 = 0.2495$, respectively. However, it should be noted that we can not achieve the minimum variance 0.2495 by the conditional plan $set(X = x + aY)$ as seen from section 3.2. Here, we assume such a conditional plan that $a$ is close to $-1/0.049$ but can not reach exactly to $-1/0.049$ (that is, $a$ satisfies $|0.049a| < 1$) under $set(X = x + aY)$. Then, this conditional plan indicates that (i) if an educational plan aimed at raising the student-faculty contact according to students' aspiration is conducted, then the average students' aspiration would also improve, but it may become smaller compared to the unconditional plan discussed above, and (ii) by raising the contact for students with lower aspiration, the variation of aspiration among students could be reduced significantly.

## 5 DISCUSSION

This paper introduced the theory of the causal effect on the variance to linear nonrecursive structural equation models. In practical studies, it is desirable to formulate a control plan based on observational data, in order to bring a response variable close to a target value with variation reduction. In order to achieve this aim, we provided methodology for evaluating the effect of a control plan on the variance of the response variable. In addition, we clarified some properties when a control plan is conducted. The results of this paper help us not only evaluate causal effects of an optimal control plan, but also understand the causal mechanism for how the variance of a response variable changes by conducting a control plan.

The approach in this paper is to introduce a new equilibrium equation, not a new dynamical equation. However, it is noted that the equations between the time indexed variables can lead to equations of the same form between the equilibrium variables (Shingaki et al. 2007). Therefore, our approach is applicable to a dynamic system as well.

**ACKNOWLEGDEMENT**

This research was partly supported by the Kurata Foundation, the Kayamori Foundation of Informational Science Advancement, the Ministry of Education, Culture, Sports, Science and Technology of Japan, and the Japan Society for the Promotion of Science.

**REFERENCES**

Ben-Israel, A. and Greville, T. N. E. (1972). *Generalized inverses: theory and applications*. John Wiley & Sons.

Bentler, P. M. and Freeman, E. H. (1983). Tests for stability in linear structural equation systems. *Psychometrika*, **48**, 143-145.

Berry, W. D. (1984). *Nonrecursive causal models*. Sage Publications.

Bollen, K. A. (1989). *Structural Equations with Latent Variables*. John Wiley & Sons.

Bowden, R. J., and Turkington, D. A. (1984). *Instrumental variables*. Cambridge University Press.

Brito, C. and Pearl, J. (2002). Generalized instrumental variables. *Proceeding of the 18th Conference on Uncertainty in Artificial Intelligence*, 85-93.

Cai, Z. and Kuroki, M. (2005). Counterfactual reasoning in linear structural equation models. *Proceeding of the 21st Conference on Uncertainty in Artificial Intelligence*, 77-84.

Duncan, O. D. (1975). *Introduction to structural equation models*. Academic Press.

Haavelmo, T. (1943). The statistical implications of a system of simultaneous equations. *Econometrica*, **11**, 1-12.

Iverson, B., Pascarella, E. T., and Terenzini, P. T. (1985). Informal faculty-student contact and commuter college freshmen. *Research in Higher Education*, **21**, 123-136.

Koster, J. T. A. (1995). Markov properties of non-recursive causal models. *Annals of Statistics*, **24**, 2148-2177.

Kuroki, M. and Cai, Z. (2004). Selection of Identifiability Criteria for Total Effects by Using Path Diagrams. *Proceedings of the 20th Conference on Uncertainty in Artificial Intelligence*, 333-340.

Kuroki, M. and Miyakawa, M. (2003). Covariate selection for estimating the causal effect of control plans using causal diagrams. *Journal of the Royal Statistical Society, Series B*, **65**, 209-222.

Kuroki, M., Miyakawa, M. and Cai, Z. (2003). Joint causal effect in linear structural equation model and its application to process analysis. *Proceeding of 9th International Workshop on Artificial Intelligence and Statistics*, 70-77.

Pearl, J. (2000). *Causality: models, reasoning, and inference*. Cambridge University Press.

Rao, C. R. (1973). *Linear statistical inference and its applications*. John Wiley & Sons.

Richardson, T. (1996a). Models of feedback: interpretation and discovery. PhD thesis. Carnegie-Mellon University.

Richardson, T. (1996b). Equivalence in non-recursive structural equation models. *Proceedings of The 11th Symposium on Computational Statistics*, 20-26.

Richardson, T. (1996c). A polynomial-time algorithm for deciding Markov equivalence of directed cyclic graphical models. *Proceedings of the 12th Conference on Uncertainty in Artificial Intelligence*, 454-461.

Richardson, T. (1996d). A discovery algorithm for directed cyclic graphs. *Proceedings of the 12th Conference on Uncertainty in Artificial Intelligence*, 462-469.

Richardson, T. (1997). A Characterization of Markov equivalence for directed cyclic graphs. *International Journal of Approximate Reasoning*, **17**, 107-162.

Richardson, T., and Spirtes, P. (1996). Automated discovery of linear feedback models. *Computation, Causation, and Discovery*, 253-303.

Shingaki, R., Kuroki, M. and Cai Z. (2007). On Causal Effects of Control Plans in Dynamic Systems. *The 6th Annual Hawaii International Conference on Statistics, Mathematics and Related Fields*, 1459.

Spirtes, P. (1995). Directed cyclic graphical representation of feedback models. *Proceedings of the 11th Conference on Uncertainty in Artificial Intelligence*, 491-498.

Strotz, R. H. and Wold, H. O. A. (1960). Recursive vs. nonrecursive systems: an attempt at synthesis. *Econometrica*, **28**, 417-427.

Tian, J. (2004). Identifying linear causal effects. *Proceedings of the 19th National Conference on Artificial Intelligence*, 104-111.